\documentclass[11pt,oneside,reqno,american]{amsart}
\usepackage[T1]{fontenc}
\usepackage[latin9]{inputenc}
\usepackage{verbatim}
\usepackage{amsbsy}
\usepackage{amstext}
\usepackage{amsthm}
\usepackage{amssymb}
\usepackage{undertilde}
\usepackage{graphicx}
\usepackage{xargs}[2008/03/08]

\makeatletter
\numberwithin{equation}{section}
\numberwithin{figure}{section}
\theoremstyle{plain}
\newtheorem{thm}{\protect\theoremname}[section]
\theoremstyle{definition}
\newtheorem{example}[thm]{\protect\examplename}
\theoremstyle{remark}
\newtheorem{rem}[thm]{\protect\remarkname}

\usepackage{mathtools}
\usepackage{amsthm}
\usepackage{pdfsync} 
\usepackage{hyperref}
\usepackage[all]{xy}

 \theoremstyle{plain}

\usepackage{fbb}
\usepackage[stix2]{newtxmath}
\usepackage[cal=boondoxo,bb=boondox,frak=boondox]{mathalfa}
\usepackage{xcolor} 
\definecolor{brown(traditional)}{rgb}{0.59, 0.29, 0.0}
\definecolor{blue(ryb)}{rgb}{0.01, 0.28, 1.0}
\definecolor{red}{rgb}{1.0, 0.0, 0.0}
\definecolor{magenta}{rgb}{1.0, 0.0, 1.0}
\definecolor{mahogany}{rgb}{0.75, 0.25, 0.0}
\definecolor{lavenderpurple}{rgb}{0.59, 0.48, 0.71}
\definecolor{olive}{rgb}{0.5, 0.5, 0.0}
\definecolor{brickred}{rgb}{0.8, 0.25, 0.33}
\definecolor{antiquefuchsia}{rgb}{0.57, 0.36, 0.51}
\definecolor{bole}{rgb}{0.47, 0.27, 0.23}
\definecolor{darkolivegreen}{rgb}{0.33, 0.42, 0.18}
\definecolor{deepjunglegreen}{rgb}{0.0, 0.29, 0.29}
\definecolor{brickred}{rgb}{0.8, 0.25, 0.33}
\definecolor{deepjunglegreen}{rgb}{0.0, 0.29, 0.29}
\definecolor{darkpastelgreen}{rgb}{0.01, 0.75, 0.24}
\definecolor{green(pigment)}{rgb}{0.0, 0.65, 0.31}
\definecolor{junglegreen}{rgb}{0.16, 0.67, 0.53}
\definecolor{officegreen}{rgb}{0.0, 0.5, 0.0}
\definecolor{seagreen}{rgb}{0.18, 0.55, 0.34}
\definecolor{teal}{rgb}{0.0, 0.5, 0.5}
\definecolor{brightgreen}{rgb}{0.4, 1.0, 0.0}
\definecolor{electricgreen}{rgb}{0.0, 1.0, 0.0}
\definecolor{malachite}{rgb}{0.04, 0.85, 0.32}

\usepackage{undertilde}
\usepackage{accents}

\makeatother

\usepackage{babel}
\providecommand{\examplename}{Example}
\providecommand{\remarkname}{Remark}
\providecommand{\theoremname}{Theorem}

\begin{document}

\global\long\def\ga{\alpha}%
\global\long\def\gb{\beta}%
\global\long\def\ggm{\gamma}%
\global\long\def\go{\omega}%
\global\long\def\gs{\sigma}%
\global\long\def\gd{\delta}%
\global\long\def\gD{\Delta}%
\global\long\def\vph{\phi}%
\global\long\def\gf{\varphi}%
\global\long\def\gk{\kappa}%
\global\long\def\gl{\lambda}%
\global\long\def\gz{\zeta}%
\global\long\def\gh{\eta}%
\global\long\def\gy{\upsilon}%
\global\long\def\gth{\theta}%
\global\long\def\gO{\Omega}%
\global\long\def\gG{\Gamma}%

\global\long\def\eps{\varepsilon}%
\global\long\def\epss#1#2{\varepsilon_{#2}^{#1}}%
\global\long\def\ep#1{\eps_{#1}}%

\global\long\def\wh#1{\widehat{#1}}%
\global\long\def\hi{\hat{\imath}}%
\global\long\def\hj{\hat{\jmath}}%
\global\long\def\hk{\hat{k}}%
\global\long\def\ol#1{\overline{#1}}%
\global\long\def\ul#1{\underline{#1}}%

\global\long\def\spec#1{\textsf{#1}}%

\global\long\def\ui{\wh{\boldsymbol{\imath}}}%
\global\long\def\uj{\wh{\boldsymbol{\jmath}}}%
\global\long\def\uk{\widehat{\boldsymbol{k}}}%

\global\long\def\uI{\widehat{\mathbf{I}}}%
\global\long\def\uJ{\widehat{\mathbf{J}}}%
\global\long\def\uK{\widehat{\mathbf{K}}}%

\global\long\def\bs#1{\boldsymbol{#1}}%
\global\long\def\vect#1{\mathbf{#1}}%
\global\long\def\bi#1{\textbf{\emph{#1}}}%

\global\long\def\uv#1{\widehat{\boldsymbol{#1}}}%
\global\long\def\cross{\times}%

\global\long\def\ddt{\frac{\dee}{\dee t}}%
\global\long\def\dbyd#1{\frac{\dee}{\dee#1}}%
\global\long\def\dby#1#2{\frac{\partial#1}{\partial#2}}%
\global\long\def\dxdt#1{\frac{\dee#1}{\dee t}}%

\global\long\def\vct#1{\bs{#1}}%

\global\long\def\partialby#1#2{\frac{\partial#1}{\partial x^{#2}}}%
\newcommandx\parder[2][usedefault, addprefix=\global, 1=]{\frac{\partial#2}{\partial#1}}%
\global\long\def\supdot{^{\!\bs{\mathord{\cdot}}}}%

\global\long\def\fall{,\quad\text{for all}\quad}%

\global\long\def\reals{\mathbb{R}}%

\global\long\def\rthree{\reals^{3}}%
\global\long\def\rsix{\reals^{6}}%
\global\long\def\rn{\reals^{n}}%
\global\long\def\eucl{\mathbb{E}}%
\global\long\def\euthree{\eucl^{3}}%
\global\long\def\euln{\eucl^{n}}%

\global\long\def\prn{\reals^{n+}}%
\global\long\def\nrn{\reals^{n-}}%
\global\long\def\cprn{\overline{\reals}^{n+}}%
\global\long\def\cnrn{\overline{\reals}^{n-}}%
\global\long\def\rt#1{\reals^{#1}}%
\global\long\def\rtw{\reals^{12}}%

\global\long\def\les{\leqslant}%
\global\long\def\ges{\geqslant}%

\global\long\def\dee{\textrm{d}}%
\global\long\def\di{d}%
\global\long\def\dX{\dee\bp}%
\global\long\def\dx{\dee x}%
\global\long\def\D{D}%

\global\long\def\from{\colon}%
\global\long\def\tto{\longrightarrow}%
\global\long\def\lmt{\longmapsto}%
\global\long\def\lhr{\lhook\joinrel\longrightarrow}%
\global\long\def\mto{\mapsto}%

\global\long\def\abs#1{\left|#1\right|}%

\global\long\def\isom{\cong}%

\global\long\def\comp{\circ}%

\global\long\def\cl#1{\overline{#1}}%

\global\long\def\fun{\varphi}%

\global\long\def\interior{\textrm{Int}\,}%
\global\long\def\inter#1{\kern0pt  #1^{\mathrm{o}}}%
\global\long\def\interior{\textrm{Int}\,}%
\global\long\def\inter#1{\kern0pt  #1^{\mathrm{o}}}%
\global\long\def\into{\mathrm{o}}%

\global\long\def\sign{\textrm{sign}\,}%
\global\long\def\sgn#1{(-1)^{#1}}%
\global\long\def\sgnp#1{(-1)^{\abs{#1}}}%

\global\long\def\du#1{#1^{*}}%

\global\long\def\tsum{{\textstyle \sum}}%
\global\long\def\lsum{{\textstyle \sum}}%

\global\long\def\dimension{\textrm{dim}\,}%

\global\long\def\esssup{\textrm{ess}\,\sup}%

\global\long\def\ess{\textrm{{ess}}}%

\global\long\def\kernel{\mathop{\textrm{\textup{Kernel}}}}%

\global\long\def\support{\mathop{\textrm{\textup{supp}}}}%

\global\long\def\image{\mathop{\textrm{\textup{Image}}}}%

\global\long\def\diver{\mathop{\textrm{\textup{div}}}}%

\global\long\def\spanv{\textrm{span}}%

\global\long\def\tr{\mathop{\textrm{\textup{tr}}}}%
\global\long\def\tran{\mathrm{tr}}%

\global\long\def\opt{\mathrm{opt}}%

\global\long\def\resto#1{|_{#1}}%
\global\long\def\incl{\mathcal{I}}%
\global\long\def\iden{\imath}%
\global\long\def\idnt{\textrm{Id}}%
\global\long\def\rest{\rho}%
\global\long\def\extnd{e_{0}}%

\global\long\def\proj{\textrm{pr}}%

\global\long\def\L#1{L\bigl(#1\bigr)}%
\global\long\def\LS#1{L_{S}\bigl(#1\bigr)}%

\global\long\def\ino#1{\int_{#1}}%

\global\long\def\half{\frac{1}{2}}%
\global\long\def\shalf{{\scriptstyle \half}}%
\global\long\def\third{\frac{1}{3}}%

\global\long\def\empt{\varnothing}%

\global\long\def\paren#1{\left(#1\right)}%
\global\long\def\bigp#1{\bigl(#1\bigr)}%
\global\long\def\biggp#1{\biggl(#1\biggr)}%
\global\long\def\Bigp#1{\Bigl(#1\Bigr)}%

\global\long\def\braces#1{\left\{  #1\right\}  }%
\global\long\def\sqbr#1{\left[#1\right]}%
\global\long\def\anglep#1{\left\langle #1\right\rangle }%

\global\long\def\bigabs#1{\bigl|#1\bigr|}%
\global\long\def\dotp#1{#1^{\centerdot}}%
\global\long\def\pdot#1{#1^{\bs{\!\cdot}}}%

\global\long\def\eq{\sim}%
\global\long\def\quot{/\!\!\eq}%
\global\long\def\by{\!/\!}%

\global\long\def\stp{\text{\small\ensuremath{\bigodot}}}%
\global\long\def\tp{\text{\small\ensuremath{\bigotimes}}}%

\global\long\def\mi#1{#1}%
\global\long\def\mii{I}%
\global\long\def\mie#1#2{#1_{1}\cdots#1_{#2}}%

\global\long\def\smi#1{\boldsymbol{#1}}%
\global\long\def\asmi#1{#1}%
\global\long\def\ordr#1{\left\langle #1\right\rangle }%

\global\long\def\symm#1{\paren{#1}}%
\global\long\def\smtr{\mathcal{S}}%

\global\long\def\perm{p}%
\global\long\def\sperm{\mathcal{P}}%

\global\long\def\oneto{1,\dots,}%

\global\long\def\lisub#1#2#3{#1_{1}#2\dots#2#1_{#3}}%

\global\long\def\lisup#1#2#3{#1^{1}#2\dots#2#1^{#3}}%

\global\long\def\lisubb#1#2#3#4{#1_{#2}#3\dots#3#1_{#4}}%

\global\long\def\lisubbc#1#2#3#4{#1_{#2}#3\cdots#3#1_{#4}}%

\global\long\def\lisubbwout#1#2#3#4#5{#1_{#2}#3\dots#3\widehat{#1}_{#5}#3\dots#3#1_{#4}}%

\global\long\def\lisubc#1#2#3{#1_{1}#2\cdots#2#1_{#3}}%

\global\long\def\lisupc#1#2#3{#1^{1}#2\cdots#2#1^{#3}}%

\global\long\def\lisupp#1#2#3#4{#1^{#2}#3\dots#3#1^{#4}}%

\global\long\def\lisuppc#1#2#3#4{#1^{#2}#3\cdots#3#1^{#4}}%

\global\long\def\lisuppwout#1#2#3#4#5#6{#1^{#2}#3#4#3\wh{#1^{#6}}#3#4#3#1^{#5}}%

\global\long\def\lisubbwout#1#2#3#4#5#6{#1_{#2}#3#4#3\wh{#1}_{#6}#3#4#3#1_{#5}}%

\global\long\def\lisubwout#1#2#3#4{#1_{1}#2\dots#2\widehat{#1}_{#4}#2\dots#2#1_{#3}}%

\global\long\def\lisupwout#1#2#3#4{#1^{1}#2\dots#2\widehat{#1^{#4}}#2\dots#2#1^{#3}}%

\global\long\def\lisubwoutc#1#2#3#4{#1_{1}#2\cdots#2\widehat{#1}_{#4}#2\cdots#2#1_{#3}}%

\global\long\def\twp#1#2#3{\dee#1^{#2}\wedge\dee#1^{#3}}%

\global\long\def\thp#1#2#3#4{\dee#1^{#2}\wedge\dee#1^{#3}\wedge\dee#1^{#4}}%

\global\long\def\fop#1#2#3#4#5{\dee#1^{#2}\wedge\dee#1^{#3}\wedge\dee#1^{#4}\wedge\dee#1^{#5}}%

\global\long\def\idots#1{#1\dots#1}%
\global\long\def\icdots#1{#1\cdots#1}%

\global\long\def\norm#1{\|#1\|}%

\global\long\def\nonh{\heartsuit}%

\global\long\def\nhn#1{\norm{#1}^{\nonh}}%

\global\long\def\bigmid{\,\bigl|\,}%

\global\long\def\trps{^{{\scriptscriptstyle \textsf{T}}}}%

\global\long\def\testfuns{\mathcal{D}}%

\global\long\def\ntil#1{\tilde{#1}{}}%

\global\long\def\pis{y}%
\global\long\def\xo{\pis_{0}}%
\global\long\def\x{x}%

\global\long\def\pib{x}%
\global\long\def\bp{X}%
\global\long\def\ii{i}%
\global\long\def\ia{\alpha}%
\global\long\def\fp{y}%
\global\long\def\piv{v}%

\global\long\def\ib{i}%
\global\long\def\is{\alpha}%

\global\long\def\pbndo{\Gamma}%
\global\long\def\bndoo{\pbndo_{0}}%
 
\global\long\def\bndot{\pbndo_{t}}%
\global\long\def\intb{\inter{\body}}%
\global\long\def\bndb{\bdry\body}%

\global\long\def\cloo{\cl{\gO}}%

\global\long\def\nor{\mathbf{n}}%
\global\long\def\Nor{\mathbf{N}}%

\global\long\def\dA{\,\dee A}%

\global\long\def\dV{\,\dee V}%

\global\long\def\eps{\varepsilon}%

\global\long\def\tv{v}%
\global\long\def\av{u}%

\global\long\def\svs{\mathcal{W}}%
\global\long\def\vs{\mathbf{V}}%
\global\long\def\avs{\mathbf{U}}%
\global\long\def\affsp{\mathcal{A}}%
\global\long\def\man{\mathcal{M}}%
\global\long\def\odman{\mathcal{N}}%
\global\long\def\subman{\mathcal{V}}%
\global\long\def\pt{p}%

\global\long\def\vbase{e}%
\global\long\def\sbase{\mathbf{e}}%
\global\long\def\msbase{\mathfrak{e}}%
\global\long\def\vect{v}%
\global\long\def\dbase{\sbase}%

\global\long\def\chart{\varphi}%
\global\long\def\Chart{\Phi}%

\global\long\def\mind{\alpha}%
\global\long\def\vb{W}%
\global\long\def\vbp{\pi}%

\global\long\def\vbt{\mathcal{E}}%
\global\long\def\fib{\vs}%
\global\long\def\vbts{W}%
\global\long\def\avb{U}%
\global\long\def\vbp{\xi}%

\global\long\def\chart{\vph}%
\global\long\def\vbchart{\Phi}%

\global\long\def\jetb#1{J^{#1}}%
\global\long\def\jet#1{j^{1}(#1)}%
\global\long\def\tjet{\tilde{\jmath}}%

\global\long\def\Jet#1{J^{1}(#1)}%

\global\long\def\jetm#1{j_{#1}}%

\global\long\def\coj{\mathfrak{d}}%

\global\long\def\alt{\mathfrak{A}}%

\global\long\def\pou{\eta}%

\global\long\def\ext{{\textstyle \bigwedge}}%
\global\long\def\forms{\Omega}%

\global\long\def\dotwedge{\dot{\mbox{\ensuremath{\wedge}}}}%

\global\long\def\vel{\theta}%

\global\long\def\Jac{\mathcal{J}}%

\global\long\def\contr{\mathbin{\raisebox{0.4pt}{\mbox{\ensuremath{\lrcorner}}}}}%
\global\long\def\fcor{\llcorner}%
\global\long\def\bcor{\lrcorner}%
\global\long\def\fcontr{\mathbin{\raisebox{0.4pt}{\mbox{\ensuremath{\llcorner}}}}}%

\global\long\def\lie{\mathcal{L}}%

\global\long\def\ssym#1#2{\ext^{#1}T^{*}#2}%

\global\long\def\sh{^{\sharp}}%

\global\long\def\nfo{\ext^{n}T^{*}\base}%
\global\long\def\dfs{\ext^{d}T^{*}\base}%
\global\long\def\dmfs{\ext^{d-1}T^{*}\base}%

\global\long\def\spc{\mathcal{S}}%
\global\long\def\sptm{\mathcal{E}}%
\global\long\def\evnt{e}%
\global\long\def\frame{\Psi}%

\global\long\def\timeman{\mathcal{T}}%
\global\long\def\zman{t}%
\global\long\def\dims{n}%
\global\long\def\m{\dims-1}%
\global\long\def\dimw{m}%

\global\long\def\wc{z}%

\global\long\def\fourv#1{\mbox{\ensuremath{\mathfrak{#1}}}}%

\global\long\def\pbform#1{\undertilde{#1}}%
\global\long\def\util#1{\raisebox{-5pt}{\ensuremath{{\scriptscriptstyle \sim}}}\!\!\!#1}%

\global\long\def\utilJ{\util J}%

\global\long\def\utilRho{\util{\rho}}%

\global\long\def\body{\mathcal{B}}%
\global\long\def\man{\mathcal{M}}%
\global\long\def\var{\mathcal{V}}%
\global\long\def\base{\mathcal{X}}%
\global\long\def\fb{\mathcal{Y}}%
\global\long\def\srfc{\mathcal{Z}}%
\global\long\def\dimb{n}%
\global\long\def\dimf{m}%
\global\long\def\afb{\mathcal{Z}}%

\global\long\def\bdry{\partial}%

\global\long\def\gO{\varOmega}%

\global\long\def\reg{\mathcal{R}}%
\global\long\def\bdrr{\bdry\reg}%

\global\long\def\bdom{\bdry\gO}%

\global\long\def\bndo{\partial\gO}%

\global\long\def\tpr{\vartheta}%

\global\long\def\mot{M}%
\global\long\def\vf{w}%
\global\long\def\const{h}%

\global\long\def\avf{u}%

\global\long\def\stn{\varepsilon}%
\global\long\def\djet{\chi}%

\global\long\def\jvf{\eps}%

\global\long\def\rig{r}%

\global\long\def\rigs{\mathcal{R}}%

\global\long\def\qrigs{\!/\!\rigs}%

\global\long\def\qd{\!/\,\!\kernel\diffop}%

\global\long\def\dis{\chi}%
\global\long\def\conf{\kappa}%
\global\long\def\invc{\hat{\conf}^{-1}}%
\global\long\def\dinvc{\hat{\conf}^{-1*}}%
\global\long\def\csp{\mathcal{Q}}%

\global\long\def\embds{\textrm{Emb}}%

\global\long\def\lc{A}%

\global\long\def\lv{\dot{A}}%
\global\long\def\alv{\dot{B}}%

\global\long\def\j{\mathop{\mathrm{j}}}%
\global\long\def\mapp{M}%
\global\long\def\J{J}%
\global\long\def\jex{\mathop{}\!\mathrm{j}}%

\global\long\def\fc{F}%
\global\long\def\load{f}%
\global\long\def\afc{g}%

\global\long\def\bfc{\mathbf{b}}%
\global\long\def\bfcc{b}%

\global\long\def\sfc{\mathbf{t}}%
\global\long\def\sfcc{t}%

\global\long\def\stm{\varsigma}%
\global\long\def\std{S}%
\global\long\def\tst{\sigma}%
\global\long\def\tstd{s}%
\global\long\def\st{\sigma}%
\global\long\def\vst{\varsigma}%
\global\long\def\vstd{S}%
\global\long\def\tstm{\sigma}%
\global\long\def\vstm{\varsigma}%

\global\long\def\stp{S_{P}}%
\global\long\def\slf{R}%

\global\long\def\crel{\Phi}%

\global\long\def\stmat{\tau}%

\global\long\def\gdiv{\bdry\textrm{iv\,}}%
\global\long\def\extjet{\mathfrak{d}}%

\global\long\def\smc#1{\mathfrak{#1}}%

\global\long\def\nhs{P}%
\global\long\def\nhsa{P}%
\global\long\def\nhsb{\underline{P}}%

\global\long\def\soc{Z}%

\global\long\def\sts{\varSigma}%
\global\long\def\spstd{\mathfrak{S}}%
\global\long\def\sptst{\mathfrak{T}}%
\global\long\def\spnhs{\mathcal{P}}%
\global\long\def\Ljj{\L{J^{1}(J^{k-1}\vb),\ext^{n}T^{*}\base}}%

\global\long\def\spsb{\text{\Large\ensuremath{\Delta}}}%

\global\long\def\ened{\mathfrak{w}}%
\global\long\def\energy{\mathfrak{W}}%

\global\long\def\ebdfc{T}%
\global\long\def\optimum{\st^{\textrm{opt}}}%
\global\long\def\scf{K}%

\global\long\def\grp{G}%
\global\long\def\gact{A}%
\global\long\def\gid{e}%
\global\long\def\gel{\ggm}%

\global\long\def\ael{\upsilon}%
\global\long\def\lal{\mathfrak{g}}%

\global\long\def\prop{P}%
\global\long\def\expr{\Pi}%

\global\long\def\aprop{Q}%

\global\long\def\flux{\omega}%
\global\long\def\aflux{\psi}%

\global\long\def\fform{\tau}%

\global\long\def\dimn{n}%

\global\long\def\sdim{{\dimn-1}}%

\global\long\def\fdens{\phi}%

\global\long\def\pform{\varsigma}%
\global\long\def\vform{\beta}%
\global\long\def\sform{\tau}%
\global\long\def\flow{J}%
\global\long\def\n{\m}%
\global\long\def\cmap{\mathfrak{t}}%
\global\long\def\vcmap{\varSigma}%

\global\long\def\mvec{\mathfrak{v}}%
\global\long\def\mveco#1{\mathfrak{#1}}%
\global\long\def\mv#1{\mathfrak{#1}}%
\global\long\def\smbase{\mathfrak{e}}%
\global\long\def\spx{\simp}%
\global\long\def\il{l}%
\global\long\def\awe{\frown}%

\global\long\def\hp{H}%
\global\long\def\ohp{h}%

\global\long\def\hps{G_{\dims-1}(T\spc)}%
\global\long\def\ohps{G_{\dims-1}^{\perp}(T\spc)}%

\global\long\def\hyper{\mathcal{S}}%

\global\long\def\hpsx{G_{\dims-1}(\tspc)}%
\global\long\def\ohpsx{G_{\dims-1}^{\perp}(\tspc)}%

\global\long\def\fbun{F}%

\global\long\def\flowm{\Phi}%

\global\long\def\tgb{T\spc}%
\global\long\def\ctgb{T^{*}\spc}%
\global\long\def\tspc{T_{\pis}\spc}%
\global\long\def\dspc{T_{\pis}^{*}\spc}%

\global\long\def\fflow{\fourv J}%
\global\long\def\fvform{\mathfrak{b}}%
\global\long\def\fsform{\mathfrak{t}}%
\global\long\def\fpform{\mathfrak{s}}%
\global\long\def\lfc{\mathfrak{F}}%

\global\long\def\maxw{\mathfrak{g}}%
\global\long\def\frdy{\mathfrak{f}}%
\global\long\def\ptnl{\psi}%
\global\long\def\tptn{\Psi}%
\global\long\def\vptn{\mathfrak{a}}%
\global\long\def\mtst{\tstd_{M}}%
\global\long\def\mvst{\vstd_{M}}%

\global\long\def\sobp#1#2{W_{#2}^{#1}}%

\global\long\def\inner#1#2{\left\langle #1,#2\right\rangle }%

\global\long\def\fields{\sobp pk(\vb)}%

\global\long\def\bodyfields{\sobp p{k_{\partial}}(\vb)}%

\global\long\def\forces{\sobp pk(\vb)^{*}}%

\global\long\def\bfields{\sobp p{k_{\partial}}(\vb\resto{\bndo})}%

\global\long\def\loadp{(\sfc,\bfc)}%

\global\long\def\strains{\lp p(\jetb k(\vb))}%

\global\long\def\stresses{\lp{p'}(\jetb k(\vb)^{*})}%

\global\long\def\diffop{D}%

\global\long\def\strainm{E}%

\global\long\def\incomps{\vbts_{\yieldf}}%

\global\long\def\devs{L^{p'}(\eta_{1}^{*})}%

\global\long\def\incompsns{L^{p}(\eta_{1})}%

\global\long\def\testf{\mathcal{D}}%
\global\long\def\dists{\mathcal{D}'}%

\global\long\def\codiv{\boldsymbol{\partial}}%

\global\long\def\currof#1{\tilde{#1}}%

\global\long\def\chn{c}%
\global\long\def\chnsp{\mathbf{C}}%

\global\long\def\current{T}%
\global\long\def\curr{R}%

\global\long\def\curd{S}%
\global\long\def\curwd#1{\wh{#1}}%
\global\long\def\curnd#1{\wh{#1}}%

\global\long\def\contrf{{\scriptstyle \smallfrown}}%

\global\long\def\prodf{{\scriptstyle \smallsmile}}%

\global\long\def\form{\omega}%

\global\long\def\dens{\rho}%

\global\long\def\simp{s}%
\global\long\def\ssimp{\Delta}%
\global\long\def\cpx{K}%

\global\long\def\cell{C}%

\global\long\def\chain{B}%
\global\long\def\A{A}%
\global\long\def\B{B}%

\global\long\def\ach{A}%

\global\long\def\coch{X}%

\global\long\def\scale{s}%

\global\long\def\fnorm#1{\norm{#1}^{\flat}}%

\global\long\def\chains{\mathcal{A}}%

\global\long\def\ivs{\boldsymbol{U}}%

\global\long\def\mvs{\boldsymbol{V}}%

\global\long\def\cvs{\boldsymbol{W}}%

\global\long\def\ndual#1{#1'}%

\global\long\def\nd{'}%

\global\long\def\cee#1{C^{#1}}%

\global\long\def\lone{\{L^{1}\}}%

\global\long\def\linf{L^{\infty}}%

\global\long\def\lp#1{L^{#1}}%

\global\long\def\ofbdo{(\bndo)}%

\global\long\def\ofclo{(\cloo)}%

\global\long\def\vono{(\gO,\rthree)}%

\global\long\def\lomu{\{L^{1,\mu}\}}%
\global\long\def\limu{L^{\infty,\mu}}%
\global\long\def\limub{\limu(\body,\rthree)}%
\global\long\def\lomub{\lomu(\body,\rthree)}%

\global\long\def\vonbdo{(\bndo,\rthree)}%
\global\long\def\vonbdoo{(\bndoo,\rthree)}%
\global\long\def\vonbdot{(\bndot,\rthree)}%

\global\long\def\vonclo{(\cl{\gO},\rthree)}%

\global\long\def\strono{(\gO,\reals^{6})}%

\global\long\def\sob{\{W_{1}^{1}\}}%

\global\long\def\sobb{\sob(\gO,\rthree)}%

\global\long\def\lob{\lone(\gO,\rthree)}%

\global\long\def\lib{\linf(\gO,\reals^{12})}%

\global\long\def\ofO{(\gO)}%

\global\long\def\oneo{{1,\gO}}%
\global\long\def\onebdo{{1,\bndo}}%
\global\long\def\info{{\infty,\gO}}%

\global\long\def\infclo{{\infty,\cloo}}%

\global\long\def\infbdo{{\infty,\bndo}}%
\global\long\def\lobdry{\lone(\bdry\gO,\rthree)}%

\global\long\def\ld{LD}%

\global\long\def\ldo{\ld\ofO}%
\global\long\def\ldoo{\ldo_{0}}%

\global\long\def\trace{\gamma}%
\global\long\def\dtrace{\delta}%
\global\long\def\gtrace{\beta}%

\global\long\def\pr{\proj_{\rigs}}%

\global\long\def\pq{\proj}%

\global\long\def\qr{\,/\,\reals}%

\global\long\def\aro{S_{1}}%
\global\long\def\art{S_{2}}%

\global\long\def\mo{m_{1}}%
\global\long\def\mt{m_{2}}%

\global\long\def\ebdfc{T}%

\global\long\def\mini{\Omega}%
\global\long\def\optimum{s^{\mathrm{opt}}}%
\global\long\def\scf{K}%
\global\long\def\opsf{\st^{\mathrm{opt}}}%
\global\long\def\doptimum{s^{\opt,{\scriptscriptstyle D}}}%
\global\long\def\loptimum{s^{\opt,{\scriptscriptstyle \mathcal{M}}}}%

\global\long\def\fsubs{M}%

\global\long\def\yieldc{B}%

\global\long\def\yieldf{Y}%

\global\long\def\trpr{\pi_{P}}%

\global\long\def\devpr{\pi_{\devsp}}%

\global\long\def\prsp{P}%

\global\long\def\devsp{D}%

\global\long\def\ynorm#1{\|#1\|_{\yieldf}}%

\global\long\def\colls{\Psi}%

\global\long\def\aro{S_{1}}%
\global\long\def\art{S_{2}}%

\global\long\def\mo{m_{1}}%
\global\long\def\mt{m_{2}}%

\global\long\def\trps{^{\mathsf{T}}}%

\global\long\def\hb{^{\mathrm{hb}}}%

\global\long\def\yieldst{s_{Y}}%

\global\long\def\yieldc{B}%

\global\long\def\lcap{C}%

\global\long\def\yieldf{Y}%

\global\long\def\sphpr{\pi_{P}}%

\global\long\def\devpr{\pi_{\devsp}}%

\global\long\def\prsp{P}%

\global\long\def\devsp{D}%

\global\long\def\ynorm#1{\|#1\|_{\yieldf}}%

\global\long\def\colls{\Psi}%

\global\long\def\cone{Q}%
\global\long\def\fpr{\Pi}%
\global\long\def\fprd{\fpr_{\devsp}}%
\global\long\def\fprp{\fpr_{\prsp}}%
\global\long\def\find{I_{\devsp}}%
\global\long\def\finp{I_{\prsp}}%
\global\long\def\fnorm#1{\norm{#1}_{\devsp}}%

\global\long\def\rig{r}%
\global\long\def\rigs{\mathcal{R}}%
\global\long\def\qrigs{\!/\!\rigs}%
\global\long\def\anv{\omega}%
\global\long\def\I{I}%
\global\long\def\mone{M_{1}}%

\global\long\def\bd{BD}%

\global\long\def\po{\proj_{0}}%
\global\long\def\normp#1{\norm{#1}'_{\ld}}%

\global\long\def\ssx{S}%

\global\long\def\smap{s}%

\global\long\def\smat{\chi}%

\global\long\def\sx{e}%

\global\long\def\snode{P}%
\global\long\def\newmacroname{}%

\global\long\def\elem{e}%

\global\long\def\nel{L}%

\global\long\def\el{l}%

\global\long\def\gr{g}%
\global\long\def\ngr{G}%

\global\long\def\eldof{\alpha}%

\global\long\def\glbs{\psi}%

\global\long\def\ipln{\phi}%

\global\long\def\ndof{D}%

\global\long\def\dof{d}%

\global\long\def\nldof{N}%

\global\long\def\ldof{n}%

\global\long\def\lvf{\chi}%

\global\long\def\amat{A}%
\global\long\def\bmat{B}%

\global\long\def\subsp{\mathcal{M}}%
\global\long\def\zerofn{Z}%

\global\long\def\snomat{E}%

\global\long\def\femat{E}%

\global\long\def\tmat{T}%

\global\long\def\fvec{f}%

\global\long\def\snsp{\mathcal{S}}%

\global\long\def\slnsp{\Phi}%
\global\long\def\dslnsp{\Phi^{{\scriptscriptstyle D}}}%

\global\long\def\ro{r_{1}}%

\global\long\def\rtwo{r_{2}}%

\global\long\def\rth{r_{3}}%

\global\long\def\fmax{M}%

\global\long\def\dform{\psi}%

\global\long\def\srfc{\mathcal{S}}%

\global\long\def\semib{\mathrm{SB}}%

\global\long\def\tm#1{\overrightarrow{#1}}%
\global\long\def\tmm#1{\underrightarrow{\overrightarrow{#1}}}%

\global\long\def\itm#1{\overleftarrow{#1}}%
\global\long\def\itmm#1{\underleftarrow{\overleftarrow{#1}}}%

\global\long\def\ptrac{\mathcal{P}}%

\global\long\def\nh#1{\hat{#1}}%
\global\long\def\nj{\hat{\jmath}}%
\global\long\def\nJ{\hat{J}}%
\global\long\def\rin#1{\mathfrak{#1}}%
\global\long\def\npi{\hat{\pi}}%
\global\long\def\rp{\rin p}%
\global\long\def\rq{\rin q}%
\global\long\def\rr{\rin r}%

\global\long\def\xty{(\base,\fb)}%
\global\long\def\xts{(\base,\spc)}%
\global\long\def\r{r}%
\global\long\def\ntm{(\reals^{n},\reals^{m})}%

\global\long\def\tproj{\frame_{\timeman}}%
\global\long\def\sproj{\frame_{\spc}}%

\global\long\def\mtn{e}%
\global\long\def\sppp{\lambda}%

\global\long\def\mtsp{\mathscr{E}}%

\global\long\def\disp{g}%
\global\long\def\diffs{G}%

\global\long\def\bv{BV}%

\title[Volumetric Growth]{Notes on Smooth and Singular Volumetric Growth}
\author{Vladimir Goldshtein$\vphantom{N^{2}}^{1}$  and Reuven Segev$\vphantom{N^{2}}^{2}$}
\address{}
\keywords{Continuum mechanics; growing bodies; volumetric growth; surface growth;
differential manifolds; de Rham currents..}
\begin{abstract}
The material structure of bodies undergoing growth is considered.
In the geometric framework of a general differential manifold modeling
the physical space and a fiber bundle modeling spacetime, body points
may be defined for any extensive property for which a smooth flux
field exists, even if the property is not conserved. Singular flux
fields are considered using the notion of a de Rham current. Writing
a generalized balance law using the boundary of the current corresponding
to a singular flux field, surface growth is unified with volumetric
growth.
\end{abstract}

\date{\today\\[2mm]
$^1$ Department of Mathematics, Ben-Gurion University of the Negev, Israel. Email: vladimir@bgu.ac.il\\
$^2$ Department of Mechanical Engineering, Ben-Gurion University of the Negev, Israel. Email: rsegev@post.bgu.ac.il}
\subjclass[2000]{70A05; 74A05.}

\maketitle

\section{Introduction}

The paper is concerned with the kinematics of growth. In particular,
we describe volumetric growth and surface growth (see \cite{Skalak1982,Taber1995})
in the same setting.

Volumetric growth is usually understood as the addition of smoothly
distributed mass to the bulk of the body. If conservation of mass
is viewed as the origin of the concept of a body in continuum mechanics,
it follows that a body undergoing volumetric growth is not a body
in the sense of continuum mechanics. It can also be argued that during
volumetric growth, body points are created due to a continuum model
of cell division. Based on \cite{SegRod-Worldlines,Segev2013,Segev_Book_2023},
we claim that volumetric growth does not contradict the notion of
traceable body points in continuum mechanics. In fact, one can associate
body points with any smoothly distributed extensive property satisfying
the Cauchy conditions for the existence of a smooth flux field. One
simply has to identify a body point with a worldline, or integral
line, of the flow field in spacetime. The existence of integral lines
follows from the existence of solutions to differential equations.
The biological notion of direct development indeed demonstrates the
existence of identifiable body points during volumetric growth.

Surface growth is usually conceived (see also \cite{SozioYavari2016,Pradhan-Yavari23}),
in distinction with volumetric growth, as the addition or removal
of body points at the boundary of a growing body. It is demonstrated
below how a weak formulation of the balance equation for an extensive
property, which allows for singular flow fields, can describe surface
growth. In other words, while a smooth distribution of the flux, under
volumetric growth, implies local conservation of body points, once
one admits non-smooth flow fields, surface growth, creation of body
points, and cell division, may occur. Thus, the two seemingly different
phenomena are unified.

The geometric setting for the formulation we propose is quite general.
Spacetime is modeled as a fiber bundle over the time axis, where the
fiber is a general $n$-dimensional manifold. In other words, the
space manifolds at two distinct instances, are diffeomorphic, but
not naturally so. We refer to this model as a \emph{proto-Galilean
spacetime}. For simplicity, the time axis is taken as $\reals$. A
smooth time-dependent flow field is represented by a differential
$n$-form on spacetime, and the source of the property is represented
by an $(n+1)$-form (note that the dimension of spacetime is $n+1$).

The generalization to the singular case is carried out using the notion
of a de Rham current\textemdash a linear functional on the space of
compactly supported infinitely smooth differential forms (or simpler
objects acting on larger spaces of less restricted differential forms).
The flux $n$-form is viewed as a special case of a one-current (a
linear functional on the space of one-forms), and the source distribution
is represented by a zero-current (a linear functional on the space
of compactly supported infinitely smooth real-valued functions). Accordingly,
the exterior derivative of the flow field that appears in the differential
balance equation, is replaced by the boundary operator on currents.

For the setting of differentiable manifolds, Section \ref{sec:SmoothExtensive}
presents the geometric representations of the various fields associated
with an extensive property. Here, a particular frame is used in spacetime
so that events may be assigned both time and place. Section \ref{sec:FluxesSpacetime}
generalizes the balance to a proto-Galilean spacetime. In particular,
the differential balance equation is written in spacetime rather than
in space. Section \ref{sec:Volumetric-growth} is concerned with body
points induced by an extensive property, even when the source of the
property does not vanish. A number of examples are presented.

In preparation for the generalization to singular fields, Section
\ref{sec:weak-form} presents the balance law in a weak integral form.
The variation that the fields are acting on is interpreted as a potential
field, so that the result of the action may be interpreted as the
rate of change of energy. Section \ref{sec:currents} reviews very
briefly the definition and elementary properties of de Rham currents,
and in particular, the boundary of a current is defined. Finally,
Section \ref{sec:Singular-growth} introduces the balance equation
in terms of de Rham currents. Finally, using a simple example, Section
\ref{sec:Surface-growth} illustrates how surface growth may be represented
using a de Rham current.

\section{\label{sec:SmoothExtensive}Smoothly distributed extensive properties
and fluxes}

We consider the balance law for an extensive property, $\expr$, in
the physical space $\spc$. For usual description of volumetric growth,
the property $\expr$ will be the mass. However, one may consider
also the electric charge, entropy, etc. In Section \ref{sec:FluxesSpacetime},
we will also write the relevant variables and balance law in spacetime
$\sptm$. 

\subsection{Basic fields}

Assuming that $\spc$ is an oriented $n$-dimensional differentiable
manifold without boundary, the density\footnote{We use the term ``density'' here in the classical physical sense,
and not in the sense of integrable densities on a manifold.} of the extensive property $\expr$ is given as a smooth $n$-differential
form, $\dens$. Thus, for a fixed control region, $\reg$, a compact
$n$-dimensional submanifold with smooth boundary of $\spc$, the
total amount of the property in $\reg$ is given by
\begin{equation}
\expr(\reg)=\ino{\reg}\dens.
\end{equation}

The evolution of the distribution of the extensive property follows
from the time-dependence of the density. That is, identifying the
time axis with $\reals$, we have a smooth mapping
\begin{equation}
\dens:\reals\times\spc\tto\ext^{n}T^{*}\spc
\end{equation}
which is a differential form for each $t\in\reals$. It follows that
for each time $t\in\reals$, 
\begin{equation}
\expr(t)(\reg)=\ino{\reg}\dens(t).
\end{equation}
Using a superimposed dot for partial differentiation relative to the
time variable, we have
\begin{equation}
\dot{\expr}(\reg)=\ino{\reg}\vform,\qquad\text{where,}\qquad\vform:=\dot{\dens}.
\end{equation}

In what follows, except for the region $\reg$, all quantities associated
with the property $\expr$ will be time-dependent, although we do
not indicate this in the notation.

It is usually assumed in continuum mechanics that the time derivative
$\dot{\expr}(\reg)$ is a result of a source of the property in space
and the total flux, $\Phi_{\bdry\reg}$, of the property that exits
$\reg$ through the boundary $\bdry\reg$. The source of the property
is given by a smooth differential $n$-form $\pform$ in $\spc$,
referred to as the \emph{source density}, so that the total production
of the property in $\reg$ is given by 
\begin{equation}
\ino{\reg}\pform.
\end{equation}

The total flux is assumed to be given in terms of a smooth differential
$(n-1)$-form, $\sform_{\reg}$ on $\bdry\reg$, to which we refer
as the \emph{flux density}, in the form
\begin{equation}
\Phi_{\bdry\reg}=\ino{\partial\reg}\sform_{\reg}.\label{eq:Tot_Flux_and_Density}
\end{equation}
Thus, the classical balance law assumes the form 
\begin{equation}
\ino{\reg}\vform=\ino{\reg}\pform-\ino{\partial\reg}\sform_{\reg}\label{eq:BalanceOnRegions}
\end{equation}
for each region $\reg$.

\subsection{The flux form and generalized Cauchy formula}

While the source density $\pform$ naturally restricts from $\spc$
to $\reg$, one does not immediately have a global field on $\spc$,
the restriction of which to $\bdry\reg$ is $\sform_{\reg}$. For
classical continuum mechanics in a Euclidean space, under Cauchy's
postulates, Cauchy's theorem asserts the existence of such a global
field, the \emph{flow field}, or the \emph{flux field.}

For continuum mechanics on manifolds, one can adapt the Cauchy postulates
to the general setting and prove the existence of a \emph{flux $(n-1)$-form},
$\flow$, on $\spc$ that induces the fields $\sform_{\reg}$ for
the various regions by restrictions. Simply put, the $(n-1)$-form
$\sform_{\reg}$ is the restriction of $\flow$ to vectors tangent
to $\bdry\reg$. Formally, the generalized Cauchy formula is
\begin{equation}
\sform_{\reg}=(T\incl_{\partial\reg})^{*}(\flow),\label{eq:CauchyFormula}
\end{equation}
where,
\begin{equation}
\incl_{\bdry\reg}:\bdry\reg\tto\spc
\end{equation}
is the natural inclusion, and $(T\incl_{\partial\reg})^{*}(\flow)$
is the pullback of $\flow$ onto $\bdry\reg$ by the tangent of the
inclusion mapping. For the general Cauchy's postulates and proof see
\cite{Segev2000,Segev_Book_2023}.

Assuming that $\flow$ is smooth, by Stokes's theorem we may now rewrite
(\ref{eq:Tot_Flux_and_Density}) as
\begin{equation}
\begin{split}\Phi_{\bdry\reg} & =\ino{\partial\reg}\sform_{\reg},\\
 & =\ino{\partial\reg}(T\incl_{\partial\reg})^{*}(\flow),\\
 & =\int_{\reg}\dee\flow.
\end{split}
\end{equation}
The balance (\ref{eq:BalanceOnRegions}) may now be written as
\begin{equation}
\ino{\reg}(\vform+\dee J)=\ino{\reg}\pform.\label{eq:BalanceOnRegions-1}
\end{equation}
Finally, since the last equation holds for an arbitrary region $\reg\subset\spc$,
one obtains the differential balance equation
\begin{equation}
\vform+\dee\flow=\pform.\label{eq:diff_balance}
\end{equation}

\subsection{Kinetic fluxes, Kinematic fluxes, and the flux bundle\label{subsec:KineticKinematicsFluxes}}

In standard continuum mechanics the Cauchy flux field is described
by a vector field, rather than an $(n-1)$-form as above. An additional
structure on spacetime, provided by a volume element, enables the
representation of a flux form, to which we will refer as a \emph{kinetic
flux}, by a vector field, to which we will refer as a \emph{kinematic
flux}.

The construction is quite straightforward. Let $\vel$ be a volume
element on $\spc$. (In fact, it is sufficient that $\vel$ be defined
on the support of $\flow$.) Then, there is a unique vector field
$v$ on $\spc$  such that
\begin{equation}
\flow=v\contr\vel,
\end{equation}
where $\bcor$ denotes the contraction of forms with vector fields.
If $\vel$ is represented locally as $\vel_{1\dots\dims}(x^{i})\lisupc{\dee x}{\wedge}n$,
and $\flow$ is represented locally by 
\begin{equation}
\sum_{k}\flow_{1\dots\wh k\dots n}(x^{i})\dee x^{1}\wedge\dots\wedge\wh{\dee x^{k}}\wedge\dots\dee x^{\dims}
\end{equation}
 (a superimposed hat representing an omission of an element), then,
$v$ is represented locally by 
\begin{equation}
v^{k}=\sgn{k-1}\frac{\flow_{1\dots\wh k\dots n}}{\vel_{1\dots n}}.\label{eq:VecInducedByFlux}
\end{equation}

As an example, if we consider the mass property, and the mass density
$\dens$ does not vanish where the flux of mass is different from
zero, a reasonable assumption, then, the components of the kinematic
flux are obtained by dividing the components of the kinetic flux by
the mass density.

It is noted that even when a volume element is not given, $\flow$
determines a unique one-dimensional subspace of the tangent space,
the \emph{flux space}, at each point where it does not vanish. One
simply chooses a volume element and computes the corresponding kinematic
flux. While, the kinematic flux depends on the choice of volume element,
by Equation (\ref{eq:VecInducedByFlux}), the subspace containing
the kinematic flux is independent of the choice of volume element.
We will refer to this one-dimensional subbundle of the tangent bundle
as the \emph{flux bundle}.

The flux bundle may also be defined by an equivalent condition. A
nonzero vector, $v$, belongs to the flux space at a point $x$ if
and only if 
\begin{equation}
v\contr\flow(x)=0.
\end{equation}
This condition also implies that the infinitesimal flux through an
$(n-1)$-dimensional infinitesimal area contained in a hyperplane
vanishes, if and only if the hyperplane contains the flux space. This
property is a generalization of the observation in the classical setting
that the infinitesimal flux through an infinitesimal area element
vanishes if the area element contains the flux vector field.

\section{Time-dependent fluxes in a proto-Galilean spacetime\label{sec:FluxesSpacetime}}

In this section, we write the balance law for a time-dependent extensive
property in an $(n+1)$-dimensional spacetime, $\sptm$, as 
\begin{equation}
\dee\fourv J=\fourv s.\label{eq:DiffBalanceEqSpTm}
\end{equation}
Here, $\fourv J$ is an infinitely smooth $n$-form, the components
of which contain the components of the time-dependent flux $\flow$
and the density $\dens$ of the extensive property. The $(n+1)$-form
$\fourv s$ includes the components of the source form $\pform$.
To motivate this formulation of the balance law, we consider below
the case where a frame is given in classical spacetime, for the case
where space is a general infinitely smooth $n$-dimensional manifold.
Note that that by ``classical spacetime'', in distinction with relativistic
mechanics, we mean that each event can be assigned a specific time.
Further details are available in \cite{Segev2013,Segev_Book_2023}.

Consequently, in accordance with classical mechanics, we assume that
an event, $\evnt$, in spacetime, $\sptm$, may be assigned a unique
time instant $t\in\reals$, where we take the time axis as $\reals$
for the sake of simplicity. Specifically, there is a projection
\begin{equation}
\frame_{\timeman}:\sptm\tto\reals.
\end{equation}

In addition, we assume that spacetime has the structure of a fiber
bundle, for which $\tproj$ is the fiber bundle projection and the
typical fiber, is an $n$-dimensional ``space'' manifold $\spc$.
We refer to this structure of spacetime as \emph{proto-Galilean} (see
\cite{Segev2022,Segev_Book_2023} for further details).

A local trivialization in spacetime is a mapping
\begin{equation}
\frame:\tproj^{-1}(U)\tto\reals\times\spc,
\end{equation}
where $U\subset\reals$ is an open subset of the time axis. Such a
trivialization is interpreted mechanically as a (local) \emph{frame
}in space time. Thus, a frame assigns to each event a time and a location
in $\spc$. We will denote the second component of $\frame$, by $\sproj$,
and so we may write $\frame=(\tproj,\sproj)$ and 
\begin{equation}
\frame(\evnt)=(\tproj,\sproj)(\evnt)=(t,x),\qquad t\in\reals,\,x\in\spc.
\end{equation}

At each event $\evnt\in\sptm$, the tangent mapping to the trivialization
is an isomorphism
\begin{equation}
T_{\evnt}\frame=(T_{\evnt}\tproj,T_{\evnt}\sproj):T_{\evnt}\sptm\tto\reals\times T_{\Phi_{\spc}(\evnt)}\spc,
\end{equation}
where it is observed that $T_{\Phi_{\timeman}(\evnt)}\reals=\reals$
and $T_{\Phi_{\timeman}(\evnt)}^{*}\reals\isom\reals$, naturally.
Denoting by $\bs 1$ the natural basis element of $T\reals$ and by
$\bs 1^{*}$ the basis element of $T^{*}\reals$, we set
\begin{equation}
\dee t=T_{\evnt}^{*}\frame_{\timeman}(\bs 1^{*})\in T_{\evnt}^{*}\sptm,\qquad\text{and}\qquad\bdry_{t}:=(T_{\evnt}\frame)^{-1}(\bs 1,0)\in T_{\evnt}\sptm.
\end{equation}
Evidently, $\dee t(\partial_{t})=1$, $T_{\evnt}\frame_{\timeman}(\partial_{t})=\bs 1$,
$\dee t(T_{\evnt}^{-1}\frame(0,v))=0$, $T_{\evnt}\frame_{\timeman}(T_{e}^{-1}\Phi(0,v))=0$,
etc.

For an alternating tensor $\go\in\ext^{r}T_{x}\spc$ we set 
\begin{equation}
\pbform{\go}:=T_{\evnt}^{*}\frame_{\spc}(\go)\in\ext^{r}T_{x}\sptm.
\end{equation}
Note that
\begin{equation}
(\bdry_{t}\contr\pbform{\go})(\lisub v,{r-1})=0
\end{equation}
Given a global frame, $\frame$, a time-dependent $r$-form, $\go$,
defined on $\spc$, induces an $(r+1)$-form, 
\begin{equation}
\fourv f=\dee t\wedge\pbform{\go},
\end{equation}
defined on spacetime. By the properties of contraction,
\begin{equation}
\bdry_{t}\contr\fourv f=\bdry_{t}\contr(\dee t\wedge\pbform{\go})=\pbform{\go}.
\end{equation}
Let $\pbform{\dens}$, $\pbform{\vform}$, $\pbform{\pform}$, and
$\pbform{\flow}$ be time-dependent differential forms on $\sptm$
corresponding, respectively, to the density, $\dens$, its rate of
change, $\vform$, the source, $\pform$, and the flux, $\flow$,
of some extensive property $\expr$ as in the previous section. Then,
we define on $\sptm$ the forms
\begin{equation}
\fourv b=\dee t\wedge\pbform{\vform},\qquad\fourv s=\dee t\wedge\pbform{\pform},\qquad\fourv J=-\dee t\wedge\pbform{\flow}+\pbform{\rho}.\label{eq:sptm-fields}
\end{equation}
Let $\flow$ and $\rho$ be represented locally as
\begin{equation}
\sum_{k}\flow_{1\dots\wh k\dots n}\lisuppwout{\dee x}1{\wedge}{\cdots}nk,\qquad\text{and}\qquad\rho_{1\dots n}\lisupc{\dee x}{\wedge}n.
\end{equation}
Then, $\pbform{\flow}$ and $\pbform{\dens}$ are represented locally
as 
\begin{equation}
\sum_{k}\flow_{1\dots\wh k\dots n}\utilde{\dee x}^{1}\wedge\cdots\wedge\utilde{\wh{\dee x{}^{k}}}\wedge\cdots\wedge\utilde{\dee x}^{n},\qquad\text{and}\qquad\rho_{1\dots n}\lisupc{\pbform{\dee x}}{\wedge}n,
\end{equation}
respectively. It follows that the local representation of $\fourv J$
is of the form
\begin{equation}
-\sum_{k}\flow_{1\dots\wh k\dots n}\dee t\wedge\utilde{\dee x}^{1}\wedge\cdots\wedge\utilde{\wh{\dee x{}^{k}}}\wedge\cdots\wedge\utilde{\dee x}^{n}+\rho_{1\dots n}\lisupc{\pbform{\dee x}}{\wedge}n.\label{eq:LocalExpFourFlux}
\end{equation}
It noted that $\fourv J$ includes the density and $\dee\fourv J$
includes the time-derivative of the density. In addition, we observe
that 
\begin{equation}
\pbform{\flow}=-\dee t\contr\fourv J.\label{eq:project-spacetime-flux}
\end{equation}

Locally, we have
\begin{equation}
\begin{split}\dee\pbform{\dens} & =\dee(\rho_{1\dots n}(t,x^{i})\lisupc{\pbform{\dee x}}{\wedge}n),\\
 & =\dot{\dens}_{1\dots n}\dee t\wedge\lisupc{\pbform{\dee x}}{\wedge}n,\\
 & =\dee t\wedge\pbform{\vform}.
\end{split}
\end{equation}
It follows that
\begin{equation}
\begin{split}\dee\fourv J & =\dee(-\dee t\wedge\pbform{\flow}+\pbform{\rho}),\\
 & =\dee t\wedge\dee\pbform{\flow}+\dee\pbform{\dens},\\
 & =\dee t\wedge(\dee\pbform{\flow}+\pbform{\vform}).
\end{split}
\end{equation}
Using Equation (\ref{eq:diff_balance}) and Equation (\ref{eq:sptm-fields}),
we conclude that
\begin{equation}
\dee\fourv J=\fourv s.\label{eq:diff-balance-sptm}
\end{equation}
While the last equation was developed using a frame, it is an invariant
equation. Hence, we accept it as the differential balance equation
for a proto-Galilean spacetime. Stokes's theorem implies immediately
that for every region $\reg\subset\sptm$,
\begin{equation}
\int_{\bdry\reg}\fourv J=\int_{\reg}\fourv s.
\end{equation}

Finally, we note that the condition $\dee\fourv s=\dee^{2}\fourv J=0$
imposes no restriction on $\fourv s$, as it is an $(n+1)$-form in
an $(n+1)$-dimensional manifold.

\section{Volumetric growth: The material structure induced by\protect \\
a smooth flux field and direct development\label{sec:Volumetric-growth}}

The existence of traceable body points is sometimes related to conservation
of mass. Thus, if the extensive property under consideration is mass,
the existence of a source term $\pform$, for the spacetime formulation,
$\fourv s$, as in the case of volumetric growth, may be considered
as an obstacle, or even a contradiction, to the introduction of body
points. In this section, we motivate the notion of body points associated
with a general extensive property, including cases where the source
term is different from zero. In particular, this holds for a source
term for the balance of mass as in the case of volumetric growth.
The existence of body points in the case of volumetric growth, offers
a mathematical description of the idealized biological notion of direct
development of an animal, so it has identifiable body points during
its development. Furthermore, one may associate conceptually, identifiable
body points with extensive properties other than mass, for example,
the heat flux field.

\subsection{Worldlines}

Let $\fourv J$ be a flux field in an $(n+1)$-dimensional spacetime,
and let $\vel$ be a volume element in $\sptm$. Then, as presented
in Section \ref{subsec:KineticKinematicsFluxes}, a kinetic flux $v$,
a vector field on $\sptm$ is induced, and it satisfies
\begin{equation}
\fourv J=v\contr\vel.
\end{equation}
The equations of Section \ref{subsec:KineticKinematicsFluxes} still
apply, where $\fourv J$ replaces $\flow$, and $\fourv s$ replaces
$\pform$.

The vector field $v$ has integral lines, trajectories 
\begin{equation}
c:(-\eps,\eps)\subset\reals\tto\sptm
\end{equation}
that solve the ordinary differential equation
\begin{equation}
\ddt c(p)=v(c(p)).
\end{equation}
Since another choice of a volume element will yield a parallel vector
field, the images of the integral lines in $\sptm$, the \emph{worldlines},
do not depend on the volume element. In fact, it can be shown that
worldlines corresponding to two different volume elements are related
by reparametrization. Hence, as one-dimensional submanifolds of $\sptm$
they depend only on $\fourv J$. The tangent space to a worldline
at any point is evidently the fiber of the flux bundle at that point.
In terms of the theory of distributions\textemdash sub-bundles of
the tangent bundle\textemdash the one-dimensional flux bundle is \emph{integrable}.

\subsection{Body points and body frames}

The notions of body points and material bodies they comprise are fundamental
in continuum mechanics. In particular, the principle of material impenetrability
implies that two body points cannot be located at the same time at
the same element of space. One may be tempted to associate the existence
of traceable body points with the conservation of mass and conclude
that in the case of volumetric growth body points are meaningless,
or that body points are necessarily added or removed from the body.

The point of view we propose is that one can associate body points
with any extensive property for which a smooth time-dependent flux
field exists. We simply identify a body point with a worldline. In
other words, we may define an equivalence class on $\sptm$ by setting
$\evnt_{1}\sim\evnt_{2}$ if $\evnt_{1}$ and $\evnt_{2}$ belong
to the same worldline. Thus, the collection of body points, the \emph{universal
body}, is the quotient set $\sptm/\sim$.

One would hope that in comparatively simple cases the universal body
will have a structure of a differentiable manifold $\spc$, and assuming
that $\fourv J$ nowhere vanishes, $\sptm$ can be identified with
the product $\reals\times\spc$. However, such a global frame does
not necessarily exist. Nevertheless, by the flow-box theorem (e.g.,
\cite[p. 245]{AbeMarsdenManifolds}), local \emph{material frames}
exist. Every event has a neighborhood the elements of which can be
parametrize by elements in an open subset of $\reals\times\rn$, where
the  coordinates in $\rn$ parametrize the body points.

\subsection{Examples}

In the examples described below, it is assumed for simplicity that
spacetime is a product bundle (trivial bundle) $\sptm=\reals^{+}\times\reals^{n}$,
where the dimensions of the space manifold will be $n=1,2$. Thus,
we will make no distinction between $\dee x^{k}$ and $\utilde{\dee x}^{k}$,
etc.
\begin{example}
\label{exa:1}For a one-dimensional space manifold, consider the flux
one-from in spacetime given by
\begin{equation}
\fourv J(t,x)=a_{t}t\dee t+a_{x}t\dee x,
\end{equation}
for constants $a_{t},\,a_{x}$.

Using Equation (\ref{eq:VecInducedByFlux}), and using the volume
element $\dee t\wedge\dee x$ (so that $\vel_{tx}=1)$, the components
of the kinematic flux are
\begin{equation}
v^{t}=a_{x}t,\qquad v^{x}=-a_{t}t.
\end{equation}
The integral lines of the vector field are illustrated in Figure \ref{fig:1}.
\begin{figure}
\begin{centering}
\includegraphics[scale=0.7]{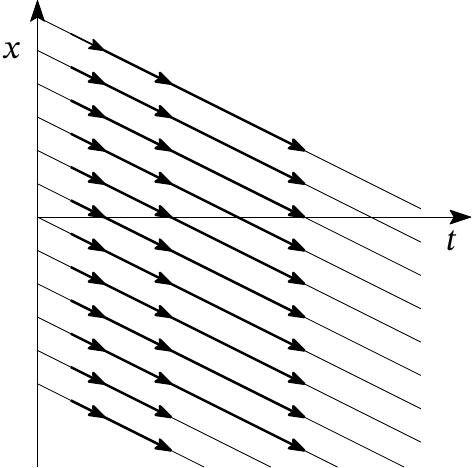}
\par\end{centering}
\caption{\label{fig:1}The kinematic flux vector field and worldlines for Example
\ref{exa:1}.}
\end{figure}
In addition,
\begin{equation}
\begin{split}\fourv s & =\dee\fourv{\fourv J,}\\
 & =a_{t}\dee t\wedge\dee t+a_{x}\dee t\wedge\dee x,\\
 & =a_{x}\dee t\wedge\dee x,
\end{split}
\end{equation}
and so
\begin{equation}
\pform=\bdry_{t}\contr\fourv s=a_{x}\dee x.
\end{equation}
Using Equation (\ref{eq:project-spacetime-flux}), the flux in the
space $\reals^{1}$ is the zero-form
\begin{equation}
\begin{split}\flow & =-\bdry_{t}\contr\fourv J,\\
 & =-\bdry_{t}\contr(a_{t}t\dee t+a_{x}t\dee x),\\
 & =-a_{t}t.
\end{split}
\end{equation}
It follows that the one form on $\reals$,
\begin{equation}
\dee\flow=\parder[x]{}(-at)\dee x=0.
\end{equation}

By the representation in Equation (\ref{eq:LocalExpFourFlux})
\begin{equation}
\dens=a_{x}t\dee x,\qquad\text{and so}\qquad\vform=\parder[t]{\dens}=a_{x}\dee x.
\end{equation}
Thus, indeed, $\vform+\dee\flow=\pform$.

Hence, although the source term is different than zero, the material
points are traceable, and they move with uniform and constant speed
$-a_{t}/a_{x}$. The exterior derivative of the flux in space vanishes,
and so all the source of the property contributes to the increase
in density.
\end{example}

\begin{example}
\label{exa:2}For a two-dimensional space manifold $\spc=\reals^{2}\setminus(0,0)$,
consider the time-dependent flux one-form in space given in polar
coordinates $(r,\ga)$ by
\begin{equation}
\flow(t,r,\ga)=a_{\ga}(t,r)\dee\ga,
\end{equation}
where for the sake of simplicity, we treat the angle $\ga$ as a global
coordinate on the circle. It is also assumed that the density two-form
is uniform relative to the natural volume element (in polar coordinates)
in the plane and constant in time. Hence,
\begin{equation}
\dens=\dens_{0}r\dee r\wedge\dee\ga,
\end{equation}
and it follows immediately that $\vform=0$.

By Equation (\ref{eq:sptm-fields}) the spacetime flux two-form is
\begin{equation}
\fourv J=-a_{\ga}(t,r)\dee t\wedge\dee\ga+\dens_{0}r\dee r\wedge\dee\ga.
\end{equation}
Noting that the natural volume element in the plane, $\vel=r\dee r\wedge\dee\ga$,
implies that $\vel_{r\ga}=r$, and using Equation (\ref{eq:VecInducedByFlux}),
we obtain for the components of the kinematic flux,
\begin{equation}
v^{t}=\dens_{0},\qquad v^{r}=\frac{a_{\ga}(t,r)}{r},\qquad v^{\ga}=0.
\end{equation}

In the particular simple case where 
\begin{equation}
a_{\ga}(t,r)=rv_{0},
\end{equation}
for a constant $v_{0}$, the kinematic flux field is 
\begin{equation}
v=\dens_{0}\bdry_{t}+v_{0}\bdry_{r}.
\end{equation}

The integral lines of the vector field are illustrated in Figure \ref{fig:2}.
\begin{figure}
\begin{centering}
\includegraphics[scale=0.7]{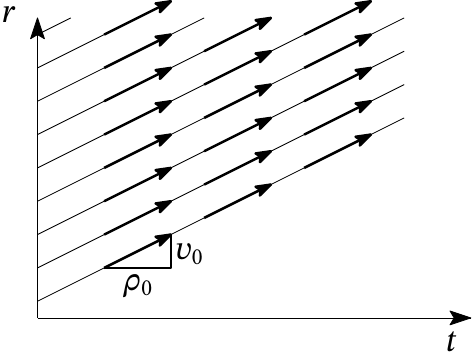}
\par\end{centering}
\caption{\label{fig:2}The kinematic flux vector field and worldlines for Example
\ref{exa:2}.}
\end{figure}
Thus, in this case, the body points move at uniform and constant velocity
$v_{0}/\dens_{0}$ in the $r$-direction. The missing point $r=0$
expands to a cavity of finite radius for $t>0$.

Since $\vform=0$, denoting partial differentiation by a comma, we
have 
\begin{equation}
\pform=\dee\flow=a_{\ga,r}\dee r\wedge\dee\ga.
\end{equation}
For the case where $a_{\ga}=rv_{0}$, 
\begin{equation}
\pform=v_{0}\dee r\wedge\dee\ga.
\end{equation}
Alternatively, 
\begin{equation}
\begin{split}\fourv s & =\dee\fourv{J,}\\
 & =-a_{\ga,r}\dee r\wedge\dee t\wedge\dee\ga,\\
 & =a_{\ga,r}\dee t\wedge\dee r\wedge\dee\ga,
\end{split}
\end{equation}
and
\begin{equation}
\pform=\bdry_{t}\contr\fourv s=a_{\ga,r}\dee r\wedge\dee\ga,
\end{equation}
as above.

We conclude that the constraint of constant and uniform density implies
that in order to compensate for the source, the body expands so that
is area increases with time.

One may also consider the case of exponential growth of a population,
as for example, the population of European rabbits in Australia since
the mid 19th century. Such a rapid growth is also typical for thermal
explosions, and the spread of pandemic viruses. When viewed as a smooth
volumetric growth, a simplified model may be given by 
\begin{equation}
a_{\ga}(t,r)=e^{r}v_{0}.
\end{equation}
Thus,
\begin{equation}
v=\dens_{0}\bdry_{t}+v_{0}\frac{e^{r}}{r}\bdry_{r}.
\end{equation}
In addition, if $\vform=0$,
\begin{equation}
\pform=\dee\flow=v_{0}e^{r}\dee r\wedge\dee\ga.
\end{equation}
The flux form in spacetime is
\begin{equation}
\fourv J=-\dee t\wedge\flow+\dens=v_{0}e^{r}\dee t\wedge\dee\ga+\dens_{0}r\dee r\wedge\dee\ga,
\end{equation}
and so,
\begin{equation}
\fourv s=\dee\fourv J=v_{0}e^{r}\dee t\wedge\dee r\wedge\dee\ga.
\end{equation}
\end{example}

\begin{example}
\label{exa:3}Using the same notation as in the previous example,
we want to illustrate the situation where no cavity of a finite radius
occurs during the volumetric growth process. Thus, it is required
that the kinematic flux vector field in spacetime and its integral
lines would be similar to those illustrated in Figure \ref{fig:3}.

\begin{figure}
\begin{centering}
\includegraphics[scale=0.7]{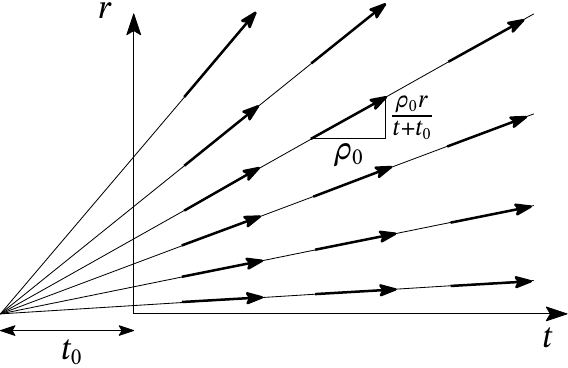}
\par\end{centering}
\caption{\label{fig:3}The kinematic flux vector field and worldlines for Example
\ref{exa:3}.}
\end{figure}
We consider, therefore, the example where,
\begin{equation}
v^{t}=\dens_{0},\qquad v^{r}=\frac{\dens_{0}r}{t+t_{0}},\qquad v^{\ga}=0.
\end{equation}

Equation (\ref{eq:VecInducedByFlux}) leads to
\begin{equation}
\fourv J=-\frac{\dens_{0}r^{2}}{t+t_{0}}\dee t\wedge\dee\ga+r\dens_{0}\dee r\wedge\dee\ga.
\end{equation}
In addition,
\begin{equation}
\fourv s=\dee\fourv J=\frac{2r\dens_{0}}{t+t_{0}}\dee t\wedge\dee r\wedge\dee\ga,
\end{equation}
and
\begin{equation}
\pform=\bdry_{t}\contr\fourv s=\frac{2r\dens_{0}}{t+t_{0}}\dee r\wedge\dee\ga=\frac{2\dens_{0}}{t+t_{0}}\vel_{\spc},
\end{equation}
where $\vel_{\spc}=r\dee r\wedge\dee\ga$ is the natural volume element
in the Euclidean plane.

Finally, the flux one-form in space is
\begin{equation}
\flow=-\bdry_{t}\contr\fourv J=\frac{\dens_{0}r^{2}}{t+t_{0}}\dee\ga.
\end{equation}
Indeed, $\vform=0$ and $\dee\flow=\pform$.

The growth process described in this simple example is reminiscent
of direct development in animals in the sense that the density is
constant in time and no cavity of finite radius is generated during
the process.
\end{example}

\section{The functional form of the balance equation\label{sec:weak-form}}

Let $\vform$, $\pform$ be the $n$-forms as above and let $\flow$
be the $(n-1)$-flux form so that in $\spc$,
\begin{equation}
\vform+\dee\flow=\pform.
\end{equation}
Let  $\gf$ be an arbitrary infinitely smooth zero-form, a real valued
function, of compact support on $\spc$. Then, multiplying the last
equation by $\gf$ and integrating over $\spc$, one obtains
\begin{equation}
\int_{\spc}\vform\wedge\gf+\int_{\spc}(\dee J)\wedge\gf=\int_{\spc}\pform\wedge\gf.\label{eq:weak1}
\end{equation}
We view the function $\gf$ as a potential function for the extensive
property, the flow of which is given by $\flow$. For example, if
the extensive property is the electric charge, then $\gf$ is the
electrostatic potential. Thus, both sides of Equation (\ref{eq:weak1})
are interpreted as the power expended during the addition of charge.

We recall that in general, for smooth $r$-form $\ga$ and $p$-form
$\ggm$, one has the identity
\begin{equation}
\dee(\ga\wedge\ggm)=(\dee\ga)\wedge\ggm+(-1)^{r}\ga\wedge\dee\ggm.\label{eq:d-wedge}
\end{equation}
Thus,
\begin{equation}
(\dee\flow)\wedge\gf=\dee(\flow\wedge\gf)-(-1)^{(n-1)}\flow\wedge\dee\gf,
\end{equation}
and Equation (\ref{eq:weak1}) assumes the form
\begin{equation}
\int_{\spc}\vform\wedge\gf+\int_{\spc}\dee(\flow\wedge\gf)=(-1)^{(n-1)}\int_{\spc}\flow\wedge\dee\gf+\int_{\spc}\pform\wedge\gf.\label{eq:weak-2}
\end{equation}

We observe that by Stokes's theorem,
\begin{equation}
\int_{\spc}\dee(\flow\wedge\gf)=\int_{\bdry\spc}\flow\wedge\gf.
\end{equation}
Thus, if $\spc$ is a manifold without boundary, so that $\bdry\spc=\varnothing$,
\begin{equation}
\int_{\spc}\dee(\flow\wedge\gf)=0,
\end{equation}
identically.

If $\reg\subset\text{\ensuremath{\spc}}$ be an infinitely smooth
$n$-dimensional submanifold with boundary, the power expended in
$\reg$ is
\begin{equation}
P_{\reg}=\int_{\reg}\vform\wedge\gf+\int_{\bdry\reg}\flow\wedge\gf=(-1)^{(n-1)}\int_{\reg}\flow\wedge\dee\gf+\int_{\reg}\pform\wedge\gf.\label{eq:weak3}
\end{equation}

Evidently, for the case of a time-dependent flux on spacetime, Equation
(\ref{eq:weak3}) assumes the form
\begin{equation}
P_{\reg}=\int_{\bdry\reg}\fourv{\flow}\wedge\gf=(-1)^{n}\int_{\reg}\fourv{\flow}\wedge\dee\gf+\int_{\reg}\fourv s\wedge\gf,\label{eq:weak4}
\end{equation}
where here, $\reg$ is an $(n+1)$-dimensional submanifold with boundary
of spacetime.
\begin{rem}
It is noted that the right hand side of the equation above is linear
in the first jet, $j^{1}\gf$, of the function $\gf$, and hence,
there is a section, $\vstd$, of the dual bundle, $J^{1}(\reg,\reals)^{*}$,
of the first jet bundle, $J^{1}(\reg,\reals)$, such that 
\begin{equation}
P_{\reg}=\int_{\reg}\vform\wedge\gf+\int_{\bdry\reg}\flow\wedge\gf=\int_{\reg}\vstd\cdot j^{1}\gf.
\end{equation}

We will not follow this point of view further in this paper, but it
is noted that this expression is in accordance with the paradigm and
results of stress theory of geometric continuum mechanics as presented,
for example, in \cite{Segev_Book_2023}.
\end{rem}

\section{Review of de Rham currents\label{sec:currents}}

De Rham currents \cite{deRham1955} provide the geometric generalization
to Schwartz distributions. While Schwartz distributions are continuous
linear functionals on the space of smooth functions with compact supports
in $\rn$, de Rham currents are continuous linear functionals on the
spaces of smooth differential forms having compact supports. The topology
on the space of forms is analogous to the space of test functions,
only one considers the convergence of the local representatives of
the forms on compact supports of the images of charts. In particular,
a de Rham $r$-\emph{current}, $T$, is a continuous linear functional
on the space of compactly supported infinitely smooth $r$-forms,
and we write $T(\go)$ for the action of the current on a form $\go$.

The following two examples illustrate the fact that currents generalize
both differential forms and manifolds.
\begin{example}
\label{exa:CurrByForm}Let $\man$ be an $n$-dimensional manifold
and let $\gf$ be an $(n-r)$-form. Then, consider the $r$-current,
$T$, defined by
\begin{equation}
T(\go)=\int_{\man}\gf\wedge\go,
\end{equation}
for every $r$-form $\go$ having a compact support in $\man$. Note
that the integral is well defined because $\gf\wedge\go$ is an $n$-form
in an $n$-dimensional manifold.
\end{example}

\begin{example}
\label{exa:CurrByManifold}Let $\man$ be an $n$-dimensional manifold,
and let $\subman$ be an $r$-dimensional submanifold of $\man$.
Then, $\subman$ induces the $r$-current, $T$, defined by
\begin{equation}
T(\go)=\int_{\subman}\go,
\end{equation}
for each $r$-form $\go$.
\end{example}

Thus, on the one hand, the current $T$ may be viewed as a generalized
$(n-r)$-form, and on the other hand, it may be thought of as a generalized
$r$-dimensional geometric object. Indeed, some fractals may be represented
by currents.

\bigskip
An $r$-current, $T$, induces an $(r-1)$-current, $\bdry T$, the
\emph{boundary} of $T$, that is defined as 
\begin{equation}
\bdry T(\psi)=T(\dee\psi)
\end{equation}
for each $(r-1)$-form $\psi$.

The boundary is a continuous operation taking $r$-currents into $(r-1)$-currents.
The two examples below demonstrate that the notion of a boundary of
a current generalizes both the exterior derivative of differential
forms and the boundary of a manifold.
\begin{example}
\label{exa:BdryCurrByForm}Let the current $T$ on the $n$-dimensional
manifold $\man$ be defined in terms of the $(n-r)$-form $\gf$ as
in Example \ref{exa:CurrByForm}. Then, for every $(r-1)$-form $\psi$
with compact support,
\begin{equation}
\begin{split}\bdry T(\psi) & =T(\dee\psi),\\
 & =\int_{\man}\gf\wedge\dee\psi,\\
 & =(-1)^{n-r}\left[\int_{\man}\dee(\gf\wedge\psi)-\int_{\man}\dee\gf\wedge\psi\right],\\
 & =(-1)^{n-r}\left[\int_{\bdry\man}\gf\wedge\psi-\int_{\man}\dee\gf\wedge\psi\right],\\
 & =(-1)^{n-r-1}\int_{\man}\dee\gf\wedge\psi,
\end{split}
\end{equation}
where in the third line above we used the identity (\ref{eq:d-wedge}),
in the fourth line we used Stokes's theorem, and in the fifth line
we used the fact that $\psi$ is compactly supported in $\man$. We
conclude that, $\bdry T$ is induced by the $(n-r+1)$-form $(-1)^{n-r-1}\dee\gf$.
\end{example}

\begin{example}
\label{exa:BdryCurrBySubman}Let $T$ be the current introduced in
Example \ref{exa:CurrByManifold}. Then, for every $(r-1)$-form $\psi$
with compact support,
\begin{equation}
\begin{split}\bdry T(\psi) & =T(\dee\psi),\\
 & =\int_{\subman}\dee\psi,\\
 & =\int_{\bdry\subman}\psi.
\end{split}
\end{equation}
It follows that the boundary of the current induced by the submanifold
$\subman$, is induce by $\bdry\subman$.
\end{example}

An additional property of currents results from the identity $\dee^{2}\psi=0$
for any differential form $\psi$. One has,
\begin{equation}
\bdry(\bdry T)(\psi)=T(\dee^{2}\psi)=0,
\end{equation}
and we conclude that 
\begin{equation}
\bdry^{2}T=0,
\end{equation}
identically. On the one hand, this property generalizes the identity
$\dee^{2}\psi=0$, for any differential form $\psi$, and on the other
hand, it generalizes the property of manifolds with boundary by which
the boundary of the boundary is empty.

\section{\label{sec:Singular-growth}Singular growth modeled by currents}

In Section \ref{sec:FluxesSpacetime} we formulated the balance differential
equation in spacetime as $\dee\fourv J=\fourv s$. Hence, for every
smooth zero-form $\gf$ with compact support in $\sptm$,
\begin{equation}
\int_{\sptm}\dee\fourv J\wedge\gf=\int_{\sptm}\fourv s\wedge\gf.
\end{equation}
Since $\dee\fourv J\wedge\gf=\dee(\fourv J\wedge\gf)-(-1)^{n}\fourv J\wedge\dee\gf$,
one has,
\begin{equation}
\begin{split}\int_{\sptm}\fourv s\wedge\gf & =\int_{\sptm}\dee(\fourv J\wedge\gf)+\int_{\sptm}(-1)^{n-1}\fourv J\wedge\dee\gf,\\
 & =\int_{\bdry\sptm}\fourv J\wedge\gf+\int_{\sptm}(-1)^{n-1}\fourv J\wedge\dee\gf,\\
 & =\int_{\sptm}(-1)^{n-1}\fourv J\wedge\dee\gf.
\end{split}
\end{equation}

Define the zero-current $S$ and the one-current $T$ by
\begin{equation}
S(\gf)=\int_{\eps}\fourv s\wedge\gf,\qquad T(\psi)=\int_{\sptm}(-1)^{n-1}\fourv J\wedge\psi,
\end{equation}
so that for any zero-form $\gf$,
\begin{equation}
\bdry T(\gf)=T(\dee\gf)=\int_{\sptm}(-1)^{n-1}\fourv J\wedge\dee\gf.
\end{equation}
Then, the currents satisfy
\begin{equation}
\bdry T=\vstd.\label{eq:Ballance current}
\end{equation}

We therefore accept Equation (\ref{eq:Ballance current}) as the balance
equation for non-smooth fluxes, where $T$ is a general one-current
representing the flux distribution in spacetime, and $\vstd$ is a
zero-current representing the source of the property.

In analogy with the smooth case, it is observed that the condition
$\bdry\vstd=\bdry^{2}T=0$, imposes no restriction on $\vstd$, as
it is a zero-current, and its boundary vanishes identically.

While in the smooth case of volumetric growth, body points, viewed
as worldlines, could not be created or destroyed, singular distributions
of fluxes enable the creation or destruction of body points.
\begin{example}
\label{exa:Ex5}\textbf{Development of an animal and reproduction.
}We present here a very simple description of the process of development
or a single animal followed by reproduction. Here, an animal is viewed
as a body point, a particle the mass of which is not necessarily conserved,
and reproduction in viewed as a splitting of the body point.

We use the same one-dimensional space manifold (the extension to three
dimensions is straightforward) and trivial spacetime bundle as in
Example \ref{exa:1}. Consider the given curves $D_{i}$, $i=\oneto3$
as in Figure \ref{fig:5}, and given differentiable functions $u_{i}(t,x)$.
Define the flux one-current

\begin{figure}[h]
\begin{centering}
\includegraphics[scale=0.7]{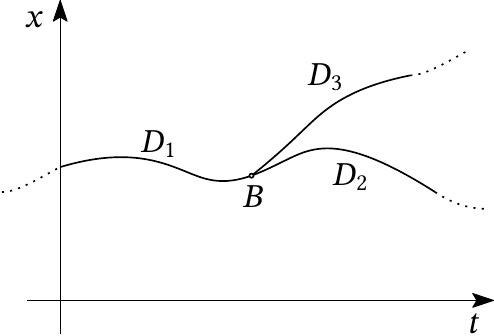}
\par\end{centering}
\caption{\label{fig:5}The worldlines for Example \ref{exa:Ex5}.}
\end{figure}

\begin{equation}
T(\go)=\sum_{i=1}^{3}\int_{D_{i}}u_{i}\wedge\go.
\end{equation}
A curve $D_{i},$represent the trajectory of a point-like animal in
space and the function $u_{i}$ may be thought of as the mass of the
animal (assumed for simplicity to depend on the current time and location
of the animal). Note that the integrals make sense as integrals of
the one-forms $u_{i}\wedge\go$ over the one-dimensional manifolds
$D_{i}$.

To compute the source current $\vstd$, we write for every smooth
zero-form $\gf$, compactly supported in $\reals^{2}$,
\begin{equation}
\begin{split}\vstd(\gf)=\bdry T(\gf) & =T(\dee\gf),\\
 & =\sum_{i=1}^{3}\int_{D_{i}}u_{i}\wedge\dee\gf,\\
 & =\sum_{i=1}^{3}\left[\int_{D_{i}}\dee(u_{i}\wedge\gf)-\int_{D_{i}}\dee u_{i}\wedge\gf\right],\\
 & =\sum_{i=1}^{3}\left[\int_{\bdry D_{i}}u_{i}\wedge\gf-\int_{D_{i}}\dee u_{i}\wedge\gf\right],
\end{split}
\end{equation}
so that
\begin{equation}
\vstd(\gf)=(u_{1}-u_{2}-u_{3})(B)\gf(B)-\sum_{i=1}^{3}\int_{D_{i}}\dee u_{i}\wedge\gf.
\end{equation}

Clearly, the one-forms $\dee u_{i}$ represent the continuous growth
of the animals, while the Dirac delta measure at $B$ of magnitude
$u_{1}-u_{2}-u_{3}$ represents the mass added, or lost, during the
reproduction event.
\end{example}

\section{Surface growth as singular volumetric growth\label{sec:Surface-growth}}

The next example illustrates how new body points can be created on
the boundary of a body in accordance with what we expect in a process
of surface growth.

Using the same notation and terminology as in the two-dimensional
space examples of Section \ref{sec:Volumetric-growth}, we consider
the product spacetime
\begin{equation}
\sptm=\{(t,r,\ga)\mid t,r\in\reals^{+},\,\ga\in S^{1}\}
\end{equation}
for which the space manifold is $\spc=\reals^{2}\setminus(0,0)$ parametrized
by polar coordinates.

\begin{figure}[h]
\begin{centering}
\includegraphics[scale=0.7]{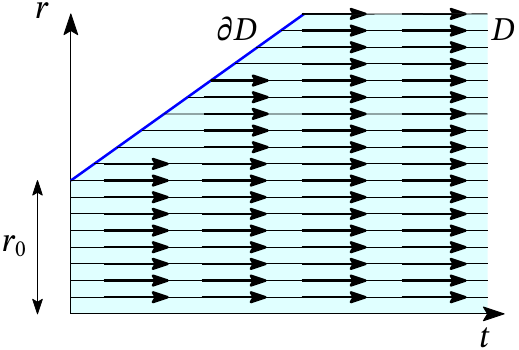}
\par\end{centering}
\caption{\label{fig:4}The kinematic flux vector field and worldlines.}
\end{figure}

Let 
\begin{equation}
D=\{(t,r,\ga)\in\sptm\mid r\les r_{0}+v_{0}t\}
\end{equation}
be the domain in spacetime, a section of which is illustrated by the
light blue shaded area in Figure \ref{fig:4}, and let 
\begin{equation}
\bdry D=\{(t,r,\ga)\in\sptm\mid r=r_{0}+v_{0}t\}
\end{equation}
represented by the blue line in the figure.

In $D$ we define the flux two-form
\begin{equation}
\fourv J=\dens_{0}r\dee r\wedge\dee\ga=\dens_{0}\vel_{\spc}.
\end{equation}
Evidently, $\fourv J$ is a smooth flux form which implies that material
point are well-defined in $D$. In fact, the kinematic flux field
corresponding to $\fourv J$ is
\begin{equation}
v=\dens_{0}\bdry_{t},
\end{equation}
as illustrated in Figure \ref{fig:4} together with the horizontal
worldlines. Furthermore, 
\begin{equation}
\fourv s=\dee\fourv J=0,
\end{equation}
everywhere in $D$, so there are no sources in $D$.

We now define a flux one-current $T$ on $\sptm$ by
\begin{equation}
T(\go)=\int_{D}\fourv J\wedge\go
\end{equation}
for every compactly supported one-form $\go$ defined on $\sptm$.
It is noted that $T$ is a current defined on $\sptm$ so it act on
forms defined in $\sptm$ (the quadrant $t-r$ in the figure) while
the integration is carried only in $D$, where $\fourv J$ is defined
and smooth.

The source zero-current, $\vstd$ is thus computed as the boundary
of $T$. Specifically, for any zero-form $\gf$ in $\sptm$
\begin{equation}
\begin{split}\vstd(\gf) & =\bdry T(\gf),\\
 & =T(\dee\gf),\\
 & =\int_{D}\fourv J\wedge\dee\gf,\\
 & =(-1)^{3-1}\left[\int_{D}\dee(\fourv{J\wedge\gf)}-\int_{D}\dee\fourv{\flow}\wedge\gf\right],\\
 & =\int_{\bdry D}\fourv J\wedge\gf.
\end{split}
\end{equation}
Thus, the source current $\vstd$ is supported on $\bdry D$.

Indeed, the example demonstrates surface growth\textemdash the creation
of body points on the evolving boundary $r=r_{0}+v_{0}t$. The fact
the the density is uniform and constant implies that the body points
remain stationary once they are created.


\begin{thebibliography}{SDM{\etalchar{+}}82}
	
	\bibitem[AMR88]{AbeMarsdenManifolds}
	R.~Abraham, J.E. Marsden, and T.~Ratiu.
	\newblock {\em Manifolds, Tensor Analysis. and Applications}.
	\newblock Springer, 1988.
	
	\bibitem[dR84]{deRham1955}
	G.~de~Rham.
	\newblock {\em Differentiable Manifolds}.
	\newblock Springer, 1984.
	
	\bibitem[PY23]{Pradhan-Yavari23}
	S.P. Pradhan and A.~Yavari.
	\newblock Accretion-ablation mechanics.
	\newblock arXiv:2307.00159v3, 2023.
	\newblock arXiv:2307.00159v3.
	
	\bibitem[SDM{\etalchar{+}}82]{Skalak1982}
	R.~Skalak, G.~Dasgupta, M.~Moss, E.~Otten, P.~Dullemeijer, and H.~Vilmann.
	\newblock Analytical description of growth.
	\newblock {\em Journal of Theoretical Biology}, 94:555--577, 1982.
	
	\bibitem[SE22]{Segev2022}
	R.~Segev and M.~Epstein.
	\newblock Proto-{G}alilean dynamics of a particle and a continuous body.
	\newblock {\em Journal of Elasticity}, 2022.
	\newblock Special issue in memory of J.~Ericksen,
	https://doi.org/10.1007/s10659-022-09929-w.
	
	\bibitem[Seg00]{Segev2000}
	R.~Segev.
	\newblock The geometry of {C}auchy's fluxes.
	\newblock {\em Archive for Rational Mechanics and Analysis}, 154:183--198,
	2000.
	
	\bibitem[Seg13]{Segev2013}
	R.~Segev.
	\newblock Notes on metric independent analysis of classical fields.
	\newblock {\em Mathematical Methods in the Applied Sciences}, 36:497--566,
	2013.
	\newblock DOI: 10.1002/mma.2610.
	
	\bibitem[Seg23]{Segev_Book_2023}
	R.~Segev.
	\newblock {\em Foundations of Geometric Continuum Mechanics}.
	\newblock Advances in Mechanics and Mathematics. Springer/Birkhauser, 2023.
	
	\bibitem[SR00]{SegRod-Worldlines}
	R.~Segev and G.~Rodnay.
	\newblock Worldlines and body points assuciated with an extensive property.
	\newblock {\em International Journal of Non-Linear Mechanics}, 38:1--9, 2000.
	
	\bibitem[SY16]{SozioYavari2016}
	Fabio Sozio and Arash Yavari.
	\newblock Nonlinear mechanics of surface growth for cylindrical and spherical
	elastic bodies.
	\newblock {\em Journal of the Mechanics and Physics of Solids}, 98, 08 2016.
	
	\bibitem[Tab95]{Taber1995}
	R.A. Taber.
	\newblock Biomechanics of growth, remodeling, and morphogenesis.
	\newblock {\em Applied Mechanics Reviews}, 48:486--545, 1995.
	
\end{thebibliography}
\newcommand{\etalchar}[1]{$^{#1}$}

\end{document}